\documentclass{easychair}
\usepackage[utf8]{inputenc}
\usepackage{graphicx}
\usepackage[frozencache]{minted}
\usepackage{tikz}
\usepackage{tikzscale}
\usetikzlibrary{positioning, calc, fit, matrix, decorations.pathreplacing,shapes.geometric}
\usepackage{adjustbox}

\usepackage{listings}
\usepackage{why3lang}

\usepackage{xspace}
\newcommand{\tez}{\textsc{Tezos}\xspace}
\newcommand{\tezla}{\textsc{Tezla}\xspace}
\newcommand{\why}{\textsc{Why3}\xspace}
\newcommand{\mich}{\textsc{Michelson}\xspace}
\newcommand{\whylson}{\textsc{WhylSon}\xspace}
\newcommand{\whyml}{\textsc{WhyML}\xspace}
\newcommand{\ocaml}{\textsc{OCaml}\xspace}

\usepackage{breakurl}

\begin{document}
\title{\whylson: Proving your \mich Smart Contracts in \why\thanks{Research Supported by the Tezos Foundation through the project FRESCO - FoRmal vErification of Smart COntracts}}

\author{Luís Pedro Arrojado da Horta\inst{1,2} \and
    João Santos Reis\inst{1,2} \and
    Mário Pereira\inst{2,4} \and %
    Simão Melo de Sousa\inst{1,2,3}
}%
\authorrunning{L.P.A. Horta et al.}
\titlerunning{\whylson: Proving your \mich Smart Contracts in \why}

\institute{Release Lab., Universidade da Beira Interior,%
    Portugal
    \and
    NOVA LINCS
    \and Cloud Computing Competence Center (C4)
    \and
    DI, FCT, Universidade Nova de Lisboa, Portugal\\
    \{luis.horta $\mid$ joao.reis $\mid$ desousa\}@ubi.pt,\ mjp.pereira@fct.unl.pt
}
\maketitle              %

\begin{abstract}

    This paper introduces \whylson, a deductive verification tool for smart
    contracts written in \mich, which is the low-level language of the Tezos blockchain. \whylson
    accepts a formally specified \mich contract and automatically translates it
    to an equivalent program written in \whyml, the programming and specification
    language of the \why framework. Smart contract instructions are mapped into a
    corresponding \whyml shallow-embedding of the their axiomatic semantics, which we
    also developed in the context of this work. One major advantage of this approach
    is that it allows an out-of-the-box integration with the \why framework, namely
    its VCGen and the backend support for several automated theorem provers. We also
    discuss the use of \whylson to automatically prove the correctness of diverse annotated smart
    contracts.

\end{abstract}

\section{Introduction} %
\label{sec:intro}
Smart contracts are reactive programs that perform general-purpose computations within a blockchain and have been used to encode arbitrarily complex business logic of digital transaction.
Since the use of smart contracts has been increasing significantly, and also since smart contracts cannot be changed once uploaded into a blockchain, it is of paramount importance to tackle the challenge of formally verifying their safety and correctness.
The main focus of our work is the formal verification of smart contracts for the Tezos blockchain \cite{tezos:2014}. Moreover, we will lean towards the \mich language and its formal specification \cite{michelson:2019}.

Our approach is to make the verification process as automatic as possible. In order to do that, we chose the deductive program verification platform \why \cite{filliatre2013why3} as the underlying proof framework tool in our smart contract verification tool.
\why is a framework aimed at automatic theorem proving through the use of external provers such as Alt-ergo \cite{bobot2013alt}, Z3 \cite{z3:DeMoura2008} or CVC4 \cite{barrett2011cvc4}. Additionally, when a proof obligation can not be automatically discharged, \why allows the user to call interactive theorem provers such as Coq or Isabelle.

This document is organised as follows: section \ref{sec:michelsonWhyML} discusses how we specified \mich language in the \why platform. Our axiomatic semantic will be described in section \ref{sec:axiom}. Section \ref{sec:translation} explains how we generate \whyml code from \mich.
On section \ref{sec:casestudies} we will details  two case studies, and 
section \ref{sec:critical} will provide the reader with a critical analysis over the work developed. 
Finally, sections \ref{sec:related} and \ref{sec:conclusion} discuss some of the related work in the field of formal verification of smart contracts and the main conclusions we gathered throughout the development of this work. 

\section{\mich Specification in \why}
\label{sec:michelsonWhyML}
\mich is a stack rewriting language for writing smart contracts for the Tezos blockchain.
For a complete explanations of the \mich language, we refer the reader to \cite{michelson:2019}.
The relevant details of this language for this work will be introduced when needed.

In the \mich language there are four primitive data types for constants, that we can name: \texttt{nat, int, string} and \texttt{bytes}. Additionally we have type \texttt{bool} for booleans and the optional type \texttt{Option $\tau$} for some value $v$ of type $\tau$, (similar to the option type in OCaml).
Some of these types are not primitive in \why and thus, we had to make some choices on how to represent them. In order to ease the correspondence between types in each language, we present the reader with table \ref{tab:types}.

\begin{table}[!ht]
\centering
\begin{tabular}{|c|c|}
\hline
\textbf{\mich Primitive Type} & \textbf{Corresponding \why type (v 1.3.0)} \\ \hline
string & string \\ \hline
nat & nat \\ \hline
int & int \\ \hline
bytes & seq bv.BV8 \\ \hline
bool & bool \\ \hline
option $\tau$ & option $\tau$ \\ \hline
unit & unit \\ \hline
\end{tabular}
\caption{Correspondence between \mich and \why types.}
\label{tab:types}
\end{table}

In \mich, both \texttt{int} (for integer constants) and \texttt{nat} (for natural number constants) have arbitrary precision, which means that computations with such constants are only limited by the Gas one is willing to pay.
When it comes to \why, type \texttt{int} already has arbitrary precision, but we had to manually define type \texttt{nat} as shown in figure \ref{lst:typeNat}. Moreover it consists of a record type with a single field \texttt{value} thus allowing us to add an invariant to type.

\begin{figure}[!ht]
\begin{why3}
  type nat = { value: int }
    invariant { value >=0 }
\end{why3}
\caption{Definition  of type \texttt{nat} in \why.}
\label{lst:typeNat}
\end{figure}

 \why supports type \texttt{string} as built-in since Version 1.3.0.
Given that a byte is a set of 8 bits, we chose to use BV8 (short for BitVector of size 8).
In \mich all data structures are immutable, and that property is still maintained with the corresponding types in \why.

In \mich, comparisons between constants of the same type are possible. Figure \ref{lst:comparableTypes} shows the definition of those comparable types in \why.

\begin{figure}[!ht]
\begin{why3}
  type comparable = 
    | Int int
    | Nat Natural.nat
    | String string
    | Bytes (seq Bytes.t)
    | Mutez int
    | Bool bool
    | Key_hash string
    | Timestamp string
    | Address string
\end{why3}
\caption{Definition of type \texttt{comparable} in \why.}
\label{lst:comparableTypes}
\end{figure}

Type \texttt{Mutez} represents \emph{micro-tez} which is in fact the smallest unit of the Tezos blockchain token. Every operation involving Mutez is mandatory checked for over/underflows. Moreover this is one of the cases where the type system really helps, because it can assure us that we do not confuse them for another numerical constant. 
The \texttt{Key\_hash} type represents the hash value of a public key.
Additionally, type \texttt{Timestamp} represents a date that can be written in a readable format according to RFC3339 \cite{rfc3339}, or in an optimised format, being the number of seconds since \emph{Epoch}.

According to the specification in \cite{michelson:2019}, comparison functions in \mich for two given constants $K_1$ and $K_2$ must return a integer value as shown in equation \ref{eq:compFunc}.

\begin{equation}
    \label{eq:compFunc}
    compare ~K_1 ~K_2 = \left\{
	\begin{array}{lll}
		-1 ~ & \mbox{se } ~K_1 < K_2  \\
		0 ~ & \mbox{se } ~K_1 = K_2  \\
		1 ~ & \mbox{se } ~K_1 > K_2  
	\end{array}
\right.
\end{equation}

In order to abide by the given specification we had to implement our own versions for said functions. As an example, we present the reader with our implementation of the comparison function for boolean constants (see figure \ref{lst:compFunc}).

\begin{figure}[!ht]
\begin{why3tiny}
let compare_bool (a b: bool) : int =
  match a, b with 
  | False,True -> (-1)
  | True,False -> 1
  | _,_ -> 0
  end
\end{why3tiny}
\caption{Comparison function for type \texttt{bool}.}
\label{lst:compFunc}
\end{figure}

\mich's execution stack contains only data or instructions, thus the type \texttt{data} is defined as depicted in figure \ref{lst:typeData}. 

\begin{figure}[!ht]
\begin{why3tiny}
type data =
  | Comparable comparable
  | Key
  | Unit
  | Some_data data
  | None_data typ
  | List (list data) typ
  | Pair data data
  | Left data typ
  | Right data typ
  | Set SetApp.set comparable_t
  | Map (my_map data) comparable_t typ
  | Big_map (my_map data) comparable_t typ
  ...
  | Mutez_Const
  | Chain_ID_Const
  | PACK_Const
  | Create_Contract_OP
  | Transfer_Tokens_OP
  | Set_Delegate_OP
  | Create_Account_OP
 with instruction = (* For convenience, all CAPITAL types are Michelson native instructions *)
  | SEQ_I instruction instruction
  ...
\end{why3tiny}
\caption{Definition of type \texttt{data} in \whyml.}
\label{lst:typeData}
\end{figure}

In order to ensure that all data is properly constructed, the predicate\hyphenation{pre-di-cate} \texttt{well\_formed\_data} is defined as shown in figure \ref{lst:predWFdt}.

\begin{figure}[!ht]
\begin{why3tiny}
predicate well_formed_data (d: data) =
    match d with
        | Map m _ _
        | Big_map m _ _ -> well_formed_map m
        | Left d _
        | Right d _
        | Some_data d -> well_formed_data d
        | Pair d1 d2 -> well_formed_data d1 /\ well_formed_data d2
        | List lst t -> well_formed_data_list lst t
        | _ -> true
    end
with well_formed_data_list (l: list data) (t: typ) =
    match l with
        | Nil -> true
        | Cons hd tl -> well_formed_data hd /\ well_formed_data_list tl t /\ typ_infer hd = t
    end
\end{why3tiny}
\caption{Definition of predicate \texttt{well\_formed\_data} in \whyml.}
\label{lst:predWFdt}
\end{figure}

For the \whyml representation of the \mich execution stack, we chose an immutable sequence, naming it type \texttt{stack\_t} and defining it as follows:\\ \mintinline{ocaml}{type stack_t = seq well_formed_data}.

Additionally we defined a function named \texttt{typ\_infer} for determining the type of a specific element in the stack.
This function gives us an extra assurance that the stack is well formed and well typed.

\section{Axiomatic Semantics in \why}
\label{sec:axiom}
In this section we present the reader with some of the more important details of our axiomatic semantics of \mich in \whyml.
Our approach is a shallow embedding of the \mich language in \whyml. Furthermore opcodes such as \texttt{SEQ} do not need
to be directly encoded given that one can take advantage of the \whyml language constructs e.g. \texttt{let $\dots$ in $\dots$} or the sequence operator ';'.

Every \mich opcode results in an abstract function in \whyml containing a set of annotations (i.e. rules).
Moreover this set of rules defines the expected behaviour of that opcode and the effect it produces on the stack.
All the opcodes take as input (at least) the stack and return a new stack.

As an example of such abstract function take the opcode \texttt{ADD} defined in \cite{michelson:2019} as the sum of the top two elements in the input stack, figure \ref{lst:addML} depicts the corresponding \whyml code.

\begin{figure}[!ht]
\begin{why3tiny}
val add (s: stack_t) (fuel: int) : stack_t
    requires { fuel > 0 }
    requires { length s >= 2 }
    requires { match typ_infer s[0].d, typ_infer s[1].d with
                | Comparable_t Int_t, Comparable_t Int_t
                | Comparable_t Int_t, Comparable_t Nat_t
                | Comparable_t Nat_t, Comparable_t Int_t
                | Comparable_t Nat_t, Comparable_t Nat_t -> true
                | _ -> false end }
    ensures  { length result = length s - 1 }
    ensures { match typ_infer s[0].d, typ_infer s[1].d with
                | Comparable_t Int_t, Comparable_t Int_t
                | Comparable_t Int_t, Comparable_t Nat_t
                | Comparable_t Nat_t, Comparable_t Int_t -> typ_infer result[0].d = Comparable_t Int_t
                | Comparable_t Nat_t, Comparable_t Nat_t -> typ_infer result[0].d = Comparable_t Nat_t
                | _ -> false end }
    ensures  { forall i: int. 1 <= i < length result -> result[i] = s[i+1] }
    ensures  { forall i: int. 1 <= i < length result -> typ_infer result[i].d = typ_infer s[i+1].d }    
    ensures  { match s[0].d, s[1].d with
                | Comparable (Int x), Comparable (Int y) ->                  
                  let res = x + y in
                  result = (mk_wf_data res) :: s[2 ..]
                | Comparable (Int x), Comparable (Nat y) ->                  
                  let res = Comparable (Int (x + (eval_nat y))) in
                  result = (mk_wf_data res) :: s[2 ..]
                | Comparable (Nat x), Comparable (Int y) ->                  
                  let res = Comparable (Int ((eval_nat x) + y)) in
                  result = (mk_wf_data res) :: s[2 ..]
                | Comparable (Nat x), Comparable (Nat y) ->                  
                  let res = Comparable (Nat (add_nat x y)) in
                  result = (mk_wf_data res) :: s[2 ..]
                | _ -> false end }
\end{why3tiny}
\caption{Definition of \texttt{ADD} in \whyml.}
\label{lst:addML}
\end{figure}

Lines 2-9 of figure \ref{lst:addML} define the pre conditions and lines 10-32 define the post conditions for this particular instruction.
Furthermore, lines 4-6 are related with the contents of the stack whereas lines 7 and 8 concern the type of elements in the stack.

\paragraph{The limit of our formalisation.}  

In the present version of the axiomatic semantics, we do not have formalised the internal details of the cryptographic operations. We have instead defined these instructions as abstract operations that follow the expected pre and post conditions.
For instance, the definition of the \texttt{sha512} instruction is shown on figure \ref{lst:shaML}.

Because the semantics of serialisation operations is not clear from the reference documentation, we also choose to abstract these operation the same way we handle cryptographic operations.

\begin{figure}[!ht]
\begin{why3tiny}
 val sha512_op (s: stack_t) (fuel: int) : stack_t
    requires { fuel > 0 }    
    requires { length s >= 1 }
    requires { typ_infer s[0].d = Comparable_t Bytes_t }
    ensures  { length result = length s }    
    ensures  { forall i: int. 1 <= i < length result -> result[i] = s[i] }
    ensures  { typ_infer result[0].d = Comparable_t Bytes_t } 
    ensures  { forall i: int. 1 <= i < length result -> typ_infer result[i].d = typ_infer s[i].d }    
    ensures  { result = (mk_wf_data Crypto_Hash_Const) :: s[1..] }
\end{why3tiny}
\caption{Definition of \texttt{sha512} in \whyml.}
\label{lst:shaML}
\end{figure}

\section{Automated Translation}
\label{sec:translation}
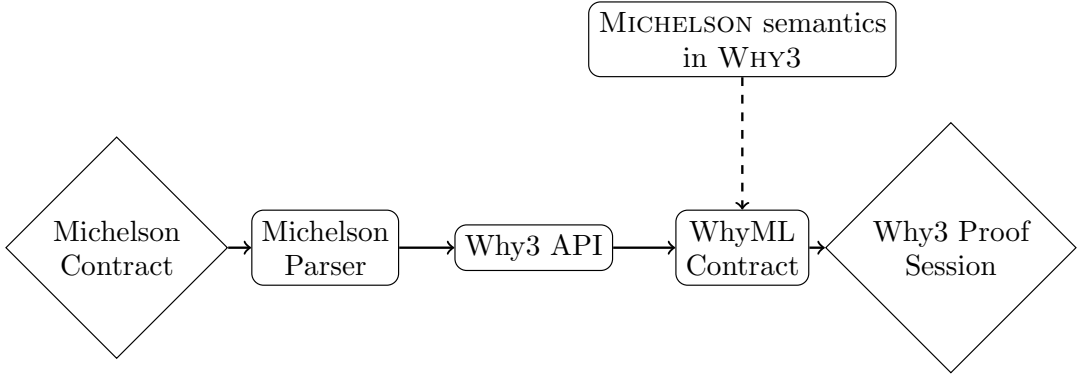
\begin{figure}[!htb]
    \centering
    \usetikzlibrary{positioning, calc, fit, matrix, decorations.pathreplacing,shapes.geometric}

\tikzstyle{io} =  [diamond, align=center, draw=black]
\tikzstyle{module} = [rectangle, rounded corners,align=center, draw=black]

\tikzstyle{arrow} = [thick,->,=>stealth]

\adjustbox{width=\textwidth}{
\begin{tikzpicture}[node distance=2.5cm]

\node (input) [io]                    {Michelson\\Contract};
\node (pr)    [module, right of=input]{Michelson\\Parser};
\node (auto)  [module, right of=pr]   {Why3 API};
\node (hck)   [module, right of=auto] {WhyML\\Contract};
\node (output)[io, right of=hck]      {Why3 Proof\\Session};
\node (sem)   [module,above of=hck]   {\mich semantics\\ in \why};

\draw [arrow]         (input) -- (pr);
\draw [arrow]         (pr)    -- (auto);
\draw [arrow]         (auto)  -- (hck);
\draw [arrow]         (hck)   -- (output);
\draw [arrow, dashed] (sem)   -- (hck);

\end{tikzpicture}
}
    \caption{Visual Structure of the Implementation.}
    \label{fig:overview}
\end{figure}

In this section we present the reader with some of the most important details about the automatic translation of the \mich written smart contract into \whyml.
For a visual representation of the WhylSon plugin structure, we refer the reader to figure \ref{fig:overview}.
In order to obtain an abstract-syntax tree of a Michelson smart contract we implemented a parser in OCaml and Menhir. This parser respects the syntax described on the Tezos documentation \cite{michelson:2019}. It allows us to obtain a data type that fully abstracts the syntax (with the exception of annotations) which we can then manipulate in order to generate the correspondent \whyml.
Furthermore, the automated translation to \whyml using the Why3 API is explained in subsection~\ref{sec:sec:api}.
Additionally, a small example of a translated \mich contract will be given in subsection \ref{sec:sec:toy}.

\subsection{Why3 API}
\label{sec:sec:api}

The core of our development is the translation of a \mich contract into an equivalent
\whyml program. Our purpose is to be able to feed the generated program to the \why
proof engine, in order to conduct formal verification on the original
contract\footnote{The correctness of the generated \whyml program implies the
correctness of the original \mich contract. At the moment, such an argument is based on
the informal reasoning that the semantics of \mich is captured by the our axiomatic
semantics developed in \why. A more rigorous rationale, which we plan to develop as
future work, must provide mathematical and/or formal evidence that \mich operational
semantics conforms to our axiomatic encoding}. It is worth highlighting that our translation
is completely done in-memory, \emph{i.e.}, \why \textit{reads} the \mich file and 
no intermediate \why file is generated in order to contain the result of translation. 
This leads to a very smooth integration with the \why framework.

The key insight of our translation mechanism is that we take the AST representation
issued by the \mich parser and, using the \why source code as an OCaml
library, we generate an AST of the \whyml language. We organise our translation code
into several mutually-recursive functions, each one dealing with the translation of a
different syntactic element of the \mich language. Consider, for instance, the
\mich instruction \whyf{ADD}. For this instruction, our parser emits an AST
containing the node \whyf{I_add}. To translate this \texttt{add} statement
into this \whyml counterpart, we build the following homomorphic translation
\begin{minted}[fontsize=\small]{ocaml}
  let rec inst = function
    | I_add -> mk_expr (Eidapp (Qident (mk_id "add"), stack_fuel_args))
    ...
\end{minted}
where \of{mk_expr} and \of{mk_id} are simply smart constructs for \whyml expressions
and identifiers, respectively. The above OCaml code creates an application expression
to our axiomatized \texttt{add} operation of Figure~\ref{lst:addML}, where the
arguments are the current stack and fuel amount. A more interesting example, and one
that shows how we take advantage of underlying translation to \whyml, is the \mich \texttt{SEQ}
operation. For such \mich statement, our parser issues node of the form \whyf{I_seq
(i1, i2)}, where~\whyf{i1} and~\whyf{i2} are the two instructions composing the
sequence. Our translation engines features the following code for this case:
\begin{minted}[fontsize=\small]{ocaml}
    | I_seq (i1, i2) ->
        mk_expr (Elet (mk_id "__stack__", false, Expr.RKnone, inst i1, inst i2))
\end{minted}
which builds the expression \whyf{let __stack__ = i1 in i2}. The Boolean constant
\texttt{false} above tells \why that this is a non-ghost expression, while the
\texttt{Expr.RKone} indicates that this is a simple locally-defined symbol, with no
direct translation to a purely-logical symbol. Let us note that the tree-like data type
produced by our translation corresponds to the AST issued by the \whyml parser, hence
no typing or name resolution information is present at this point.

Having defined our \mich to \whyml transformation function, we want to integrate it
with the \why framework in a completely transparent fashion for the end user. This
means that we want to use the \why proof engine over a \mich contract, as if this was
the native language of the framework. One can easily extend the \why framework with the
support for new input languages via its plugin machinery. This is as simple as providing
a parser and a translation function from the said language into one of the \whyml
internal AST. Finally, in order to register the newly-developed plugin into the \why
configuration base, one simply states the extension of files that should be processed by 
the devised translation. In our particular case, we write the following:
\begin{minted}[fontsize=\small]{ocaml}
  let () =
    Env.register_format mlw_language "michelson" ["tz"] read_channel
      ~desc:"Michelson format"
\end{minted}
Here, \texttt{mlw\_language} indicates that the target of our translation is a \whyml
program and \texttt{read\_channel} is the function that calls the \mich parser and
feeds the produced AST to our transformation mechanism. With all this machinery in
place, one can then call \why directly on a \texttt{.tz} file. For instance, if one
wishes to formally verify the contract contained in file \texttt{foo.tz}, using our
plugin, it is just a matter of doing
\begin{verbatim}
  $ why3 ide foo.tz
\end{verbatim}
which would open the \why graphical Integrated Development Environment over the result
of the \mich contract translation.

\subsection{A Trivial Example}
\label{sec:sec:toy}
For a better understanding of this automated translation, we present the reader with a visual toy example.
Consider the \mich contract depicted in figure \ref{lst:addnat-michelson}.

\begin{figure}[!ht]
\begin{lstlisting}
parameter nat;
storage nat;
code { UNPAIR; ADD;    
       NIL operation; PAIR };
\end{lstlisting}
\caption{Toy example of a  \mich contract.}
\label{lst:addnat-michelson}
\end{figure}

This is a very simple contract, in fact it takes the \texttt{nat} it received as parameter and adds it to the \texttt{nat} in the storage. Basically it just adds two natural numbers.
As shown by the figure above, the \mich contract does not contain any pre or post conditions, but WhylSon is able to directly infer four safety conditions, namely about the length and type of both the input and the output stacks.
The \whyml code generated by WhylSon is shown in figure \ref{lst:addnat-WhyML}.

\begin{figure}[!ht]
\begin{why3tiny}
use axiomatic.AxiomaticSem
use dataTypes.DataTypes
use seq.Seq
use int.Int
let test (__stack__: stack_t) (__fuel__: int) : stack_t
  requires { (length __stack__) = 1 }
  requires { __fuel__ > 0 }
  requires { (typ_infer (d (__stack__[0])))
               = (Pair_t (Comparable_t Nat_t ) (Comparable_t Nat_t )) }
  ensures { (length result) = 1 }
  ensures { (typ_infer (d (result[0])))
              = (Pair_t (List_t Operation_t ) (Comparable_t Nat_t )) } =
  let __stack__ =
    let __stack__ = unpair __stack__ __fuel__ in
    (let __stack__ = add __stack__ __fuel__ in
     (let __stack__ = nil_op __stack__ __fuel__ Operation_t  in
      (pair __stack__ __fuel__))) in
  __stack__
\end{why3tiny}
\caption{The \whyml generated code for the toy example.}
\label{lst:addnat-WhyML}
\end{figure}

Using only the \texttt{split\_vc} transformation this example generates 26 verification conditions that are quickly dispatched by Alt-ergo\cite{bobot2013alt}.

\section{Case Studies}
\label{sec:casestudies}
In this section we discuss two case studies, namely the multisig and factorial smart contracts and explain how safety and functional correctness can be proved within \whylson.
For the sake of brevity we will elaborate on the details the proof of safety for the multisig smart contract and on the functional correctness of the factorial smart contracts.
We will detail the proof of correctness of the factorial smart contracts since this contract highlight particularly well our purpose to show the advantages but also the drawbacks of our approach.

Both of these contracts were manually translated to WhyML.
The complete details of the formalisation and the proof of these smart contracts can be found at \burl{https://gitlab.com/releaselab/fresco/whylson}

\subsection{Multisig}
There are several versions of the multisig contract, and the one we used can be found in \cite{multisig:2020}.
We separated the multisig contract into three parts.
The first one is the majority of the contract, the second one is the loop which iterates over the list of keys and optional signatures
(\texttt{iter\_multisig}) and finally, the third part (\texttt{outer\_if\_left}) is where the operation requested by the signers is produced.
For the sake of brevity, these fuctions are not depicted here.

Figure \ref{lst:multisig-head-whyml} contains the part of the \texttt{multisig} function that represents the contract.
\begin{figure}[!ht]
  \begin{why3tiny}
    let multisig_contract (in_stack: stack_t) (fuel: int) : stack_t
    requires { fuel > 0 }
    requires { length in_stack = 1 }
    requires { typ_infer in_stack[0].d = Pair_t parameter storage }
    ensures  { length result = 1 }
    ensures  { typ_infer result[0].d = Pair_t (List_t Operation_t) storage }
    raises { Failing }
    =
    let s = unpair in_stack fuel in
    let s = swap s fuel in
    let s = dup s fuel in
    ...
    let s = iter_multisig s fuel
    ensures {
        typ_infer result[0].d = Comparable_t Nat_t /\ (* @ valid *)
        typ_infer result[1].d = List_t (Option_t Signature_t) /\
        typ_infer result[2].d = Comparable_t Bytes_t /\
        typ_infer result[3].d = Or_t
        (Pair_t (Comparable_t Mutez_t) (Contract_t Unit_t))
        (Or_t
        (Option_t (Comparable_t Key_hash_t))
        (Pair_t (Comparable_t Nat_t) (List_t Key_t))) /\
        typ_infer result[4].d = storage
      } in
    ...
  \end{why3tiny}
  \caption{Part of the multisig contract in \whyml.}
  \label{lst:multisig-head-whyml}
\end{figure}
We ask the reader to notice that it has only two pre conditions and two post conditions regarding the size and the type of the stack.
In order to prove the last post condition, we had to equip some of the code in the contract with some additional typing information.
An example of such typing information is the one below the \texttt{iter\_multisig} line.
Furthermore, this complement was necessary to help the SMTs check some of the pre conditions needed for the instructions in the middle.

This code generated a total of 758 VCs, 750 of which were dispatched by Alt-ergo \cite{bobot2013alt}, Z3 \cite{z3:DeMoura2008} and CVC4\cite{barrett2011cvc4} dispatched 4 verification conditions each.

\subsection{Factorial}
The contract depicted in figure \ref{lst:fact-michelson} is the \mich version of the factorial calculation.
This contract calculates the factorial of a given natural number interactively.
The contract receives as parameter the number whose factorial is going to be calculated and stores the result in the storage.
It starts by dropping the previous storage and pushes an initial accumulator and iterator as the value 1. Then it compares the parameter value with 0 and if it's different, it enters the loop to calculate the factorial.

\begin{figure}[!ht]
  \begin{lstlisting}[basicstyle=\scriptsize]
parameter nat;
storage nat;
code {  CAR; PUSH @index nat 1; DUP @acc;
        DIP 2 { DUP; PUSH nat 0; COMPARE; NEQ };
        DIG 2;
        LOOP { DIP { DUP;
                     DIP { PUSH nat 1; ADD @ipp } };
               MUL;
               DIP { DIP { DUP };
                     DUP;
                     DIP { SWAP }; 
                     COMPARE; LE };
               SWAP };
        DIP { DROP; DROP };
        NIL operation; PAIR };
\end{lstlisting}
  \caption{Factorial \mich contract.}
  \label{lst:fact-michelson}
\end{figure}

Inside the loop \emph{body}, the stack has size three, where the top element is the temporary result, the middle element is the index of the iteration and the bottom element is the input parameter.
Since the \emph{body} of the loop is where the computation actually happens, we will focus on the respective portion of \whyml code depicted in figure \ref{lst:fact-whyml}.
For shortness we omitted typing information in between instructions as well as size and length pre and post conditions.
The only specification that we left was the one regarding functional correctness.

\begin{figure}[!ht]
  \begin{why3tiny}
    let loop_body (s: stack_t) (fuel: int) : stack_t
    requires { match s[0].d,s[1].d with
    | Comparable(Nat res),Comparable(Nat n) -> fact (n.value - 1) = res.value
    | _ -> false end }
    ensures  { match s[0].d,s[1].d with
    | Comparable (Nat res_old), Comparable (Nat n_old)->
    fact n_old.value = n_old.value  * res_old.value
    | _ -> false
    end }
    ensures  { match s[1].d, result[1].d with
        | Comparable (Nat i), Comparable (Nat b) ->  fact i.value  =  b.value
        | _ -> false end }
    =
    let s =
    let top = s[0] in let s = s[1..] in (* DIP *)
    let s = dup s fuel in
    let s =
    let top = s[0] in let s = s[1..] in (* DIP *)
    let s = push s fuel (mk_wf_data (Comparable (Nat (to_nat 1)))) in
    let s = add s fuel in
    push s fuel top in
    push s fuel top in
    let s = mul s fuel in
    let s =
    let top = s[0] in let s = s[1..] in (* DIP *)
    let s =
    let top = s[0] in let s = s[1..] in (* DIP *)
    let s = dup s fuel in
    push s fuel top in
    let s = dup s fuel in
    let s =
    let top = s[0] in let s = s[1..] in  (* DIP *)
    let s = swap s fuel in
    push s fuel top in
    let s = compare_op s fuel in
    let s = le s fuel in
    push s fuel top in
    swap s fuel
  \end{why3tiny}
  \caption{Factorial \whyml contract.}
  \label{lst:fact-whyml}
\end{figure}
The first pre condition assures us that the value stored at the top of the input stack is in fact the value of factorial up to the previous iterations.
The last post condition ensures that the value stored at the top of the result stack is the value of factorial up to the current iteration.
This code generated 2890 VCs, of which 2671 were dispatched by Alt-ergo\cite{bobot2013alt}, and the remaining 219 by Z3\cite{z3:DeMoura2008}.

\section{Critical Analysis and Future Work}
\label{sec:critical}
As stated in the previous sections, we chose \why as the main tool for our approach at verifying \tez smart contracts
based on one simple goal, that was to automate as much as possible the verification effort on the user side. Despite this being a clear and well defined objective, we came across some adversities which will be explained in the remainder of this section.

As shown in subsection \ref{sec:michelsonWhyML} we defined numerous algebraic data types to reflect the grammar of the \mich language in a direct correspondence.
This decision has proven itself to have consequences, since SMTs find them very hard to work with. %
One possible solution is to go even further in our shallow embedding and try to map as much as possible every \mich type directly into \whyml native types. Moreover this would allow us to remove some algebraic constructors from our definition.
We first came across with this issue when trying to prove safety properties of longer contracts, as in the case of multisig.
Somewhere in the middle of the contract one could notice that the SMTs were struggling to prove some pre and post conditions
regarding the type that the instructions were expecting. In an attempt to minimise their effort, we decided to propagate typing information throughout the contract.
This was a very time consuming process, because, even if the task is systematic, we had to do it manually. This indicates that there is a clear room for automation here.
As a future improvement one could write an interpreter that would work alongside the translation mechanism and automatically propagate such conditions throughout the generated \whyml code.

When it comes to functional correctness of a \mich written contract, it is far from simple for one to infer what a contract does just by looking at it.
Since \whylson is not parsing specification from \mich contracts yet, %
we had to add it manually to the contracts we tested.
When proving the functional correctness of the factorial contract, we noticed that almost every proof needed numerous assertions in the middle of the \whyml code.
Additionally, the proof trees for these goals were far too long, but were almost entirely based on hypothesis rewriting.
Going forward we think that we might adopt some sort of proof by reflection mechanism \cite{bertotProofReflection2004,melquiond:hal-01699754} so that this proving process becomes less tedious.

Finally the fact that \mich is a stack rewriting language makes us operate solely over one data structure with no clear separation between values and instructions.
This also increases the struggle that SMTs have when it comes to guaranteeing some frame conditions.
With this thought in mind, we are considering adopting a higher level language such as Albert \cite{albert} or an intermediate representation \tezla \cite{tezla}.
On one hand, if we choose to go with Albert, we would use it as the input language to \whylson and then using the \why code extraction mechanism described in \cite{teseMario:2018} one could extract the \mich certified code. Furthermore this last effort only amounts to writing a new printer that translates the internal \why AST into compilable \mich code.
On the other hand if we decide to go with \tezla the input language stays the same (i.e. \mich) but the \whyml generated code would be based on \tezla which we think would facilitate some of the proofs.

\section{Related Work}
\label{sec:related}
When it comes to formal verification of smart contracts, there are some efforts towards the design of verification platforms for said contracts.
For instance, the work of \emph{Nehai, Z. e Bobot, F.} presented in \cite{ethwhy:2019} where they use Why3 to write smart contracts for the Ethereum blockchain \cite{eth:buterin2014next}. 
Also \emph{Bhargavan, K., et al.} developed a framework for analysis and verification of functional correctness of Ethereum (ETH) smart contracts by translation into $F^\ast$ \cite{Bhargavan:2016}.
Moreover, for the same blockchain, \emph{Abdellatif, T.} and \emph{Brousmiche, K.} used the BIP framework \cite{basu2006modeling} for modeling and verifying said contracts using statistical model checking.
Using the Coq Proof Assistant, \emph{Zheng Yang} and \emph{Hang Lei} combined symbolic execution with higher order theorem proving into a tool called FEther aimed at verifying Ethereum smart contracts \cite{FEther:2018}.
The CertiK company has developed a commercial framework for formally verifying smart contracts and blockchain ecosystems \cite{certik}.
\emph{Marvidou, A.} and \emph{Laska, A.} presented FSolidM in \cite{mavridou2017designing}, a framework that allows its users to write more secure contracts for ETH using a graphical interface for designing finite state machines that will then be automatically translated into ETH smart contracts.
In \cite{sergey2018scilla}, \emph{Sergey, I., et al.} describe SCILLA, an intermediate language for ETH smart contracts that is amenable to formal verification.

When it comes to \mich formalisation, \emph{Bernardo, B., et al.} specified a big-step semantic for Michelson using the Coq proof assistant \cite{bernardo2019mi} that serves as a base for a verification framework. This work differs from ours because we focus on the automation of the verification process. This fact relies on Why3 where the proof obligations are dispatched to external provers, where as in Coq the proof is made manually.
The Archetype language \cite{archetype:2020} is a domain specific language that allows for formal specification of Tezos smart contracts, %
which in turn are translated to \whyml for use in \why, as a back-end. In this work, the contracts to be verified are Archetype contracts.
Moreover, we chose \mich as the object of our verification process, thus mitigating the need for the smart contract writer to learn yet another language for smart contract development.

\section{Conclusions}
\label{sec:conclusion}
In this paper we presented \whylson, a tool for automated formal verification of \mich smart contracts.
Moreover, \whylson is the result of several implementations also described in this document, namely
a \mich parser, an axiomatic semantics and a translation function, all leading to a shallow embedding of \mich in \whyml.

The first steps to the automatic proof in \why of manually annotated \mich smart contracts were done,
since we were able to use our axiomatic semantics and our \why plugin to successfully 
write and prove several \mich contracts. 
Furthermore, the plugin development proved itself simple, due to the fact that the \why platform exposes its API as an \ocaml library.

In practice we found that some of the proof trees were bigger than expected and required user intervention, thus threatening our main purpose of automation. 
We are aware that this is a consequence of our encoding of \mich types as tree-like data structures. %
Our perspective is that using one or more of the solutions discussed in \ref{sec:critical} we can mitigate this issue, leading us to a platform that
allows the user to conduct formal verification of \mich written smart contract with an elevated degree of automation.

As a final thought, we think that proving \mich contracts has a certain advantage over proving some other formulation, since what is effectively executed is the \mich smart contract and also because this approach can be used a back-end in developing reliable smart contract in any higher level language such a LIGO\cite{ligo} or SmartPy\cite{smartpy}. 

Nevertheless, smart contracts developers will implement their smart contracts in a higher level language than \mich. In this setting, it is also relevant to be able to formally prove these smart contracts at a level that programmers understand and be involved with. 
So an interesting long-term line of work to explore is to connect \whylson with certifying-certified compilation techniques and platforms. For instance, we should evaluate how the integration of \whylson with, \emph{e.g.}, the Archetype platform \cite{archetype:2020}, that also makes use of \why, can benefit the automatic proof of \mich smart contracts. We should also evaluate how \whylson could benefit from  rigorously designed compilers as the one designed for Albert \cite{albert} to Mi-cho-coq \cite{bernardo2019mi}. For the \why platform, such an endeavour could make use of the techniques introduced by \emph{Clochard, M., et al.} in \cite{clochardDeductiveVerificationGhost2020}.

\bibliographystyle{abbrvurl}
\bibliography{references.bib}

\end{document}